\documentclass{article}
\usepackage{spconf,cite}
\usepackage{amsmath,graphicx,amsfonts,epsfig}
\usepackage{times}
\usepackage{amsbsy}
\usepackage{latexsym}
\usepackage{amssymb}
\usepackage{color}
\usepackage{algorithm}
\usepackage{algorithmic}
\usepackage{enumerate}

\title{Joint SIC and
Relay Selection for Cooperative DS-CDMA Systems} \vspace{-1cm}

\name{ Jiaqi Gu{$^*$} and Rodrigo C. de Lamare{$^{*\dag}$}
\vspace{-0.5em}}
\address{Communications Research Group, Department of Electronics, University of York, U.K.{$^*$}\\
 CETUC, PUC-Rio, Rio de Janeiro, Brazil{$^\dag$}\\
Emails: jg849@york.ac.uk, rodrigo.delamare@york.ac.uk
\vspace{-0.25em}}

\begin{document}
\linespread{0.975} \setlength{\abovedisplayskip}{0.4mm}
\setlength{\belowdisplayskip}{0.4mm} \setlength\floatsep{8pt}
\setlength\textfloatsep{8pt} \setlength\intextsep{7.5pt}
\setlength{\abovecaptionskip}{2pt}
\setlength{\belowcaptionskip}{2pt}

\maketitle
\let\thefootnote\relax\footnote{This work is funded by the ESII consortium under task 26 for
low-cost wireless ad hoc and sensor networks}\vspace{-2em}

\begin{abstract}
In this work, we propose a cross-layer design strategy based on
a joint successive interference cancellation (SIC) detection
technique and a multi-relay selection algorithm for the uplink of
cooperative direct-sequence code-division multiple access (DS-CDMA)
systems. We devise a low-cost greedy list-based SIC (GL-SIC) strategy with RAKE receivers as
the front-end that can approach the maximum likelihood detector
performance. %Unlike prior art, the proposed GL-SIC algorithm
%exploits the Euclidean distance between users of interest, multiple
%ordering and their constellation points to build an effective list
%of detection candidates.
We also present a low-complexity multi-relay selection algorithm based on greedy techniques that can
approach the performance of an exhaustive search. %A cross-layer
%design strategy that brings together the proposed GL-SIC algorithm
%and the greedy relay selection is then developed.
Simulations show an excellent bit error rate performance of the proposed
detection and relay selection algorithms as compared to existing
techniques.
\end{abstract}


\vspace{-0.4cm}
\section{Introduction}
\vspace{-0.35cm}
In wireless communications, fading induced by multipath propagation
has a detrimental effect on the received signals. Indeed, several fading can lead to a
degradation of the transmission of information and the reliability
of the network. In order to mitigate this effect, modern diversity
techniques like cooperative diversity have been widely considered in
recent years \cite{Proakis}. %With cooperative diversity, a message
%can be transmitted via independent channels and after being
%processed at the relay, the information is forwarded to the
%destination.
Several cooperative schemes have been proposed %in the literature
 \cite{sendonaris,Venturino,laneman04} and among the most
effective ones are Amplify-and-Forward (AF) and Decode-and-Forward
(DF) \cite{laneman04,TDS_RS,TDS,smce,armo}. For an AF protocol,
relays cooperate and amplify the received signals with a given
transmit power, this method is simple except that the relays amplify
their own noise. With DF protocol, relays decode the received
signals and then forward the re-encoded message to the destination.
Consequently, better performance and lower power consumption can be
obtained when appropriate decoding and relay selection strategies
are applied.

DS-CDMA systems are a multiple access technique that can be
incorporated with cooperative systems in ad hoc and sensor networks
\cite{Bai,Souryal,Levorato,sit}. Due to the multiple access
interference (MAI) effect that arises from nonorthogonal received
waveforms, the system is adversely affected. To address this issue,
multiuser detection (MUD) techniques have been developed in
\cite{Verdu1} as an effective approach to suppress MAI. The optimal
detector, known as maximum likelihood (ML) detector, has been
proposed in \cite{Verdu2}. However, this method is infeasible for ad
hoc and sensor networks considering its computational complexity.
Motivated by this fact, several sub-optimal strategies have been
developed: linear detectors \cite{Lupas}, successive interference
cancellation (SIC) \cite{Patel,bsic,mbsic,Li1}, parallel
interference cancellation (PIC) \cite{Varanasi,itic,mfpic} and
decision feedback detectors \cite{mber,RCDL1,stmf,mfdf,dfjio,mbdf}.

Prior studies on relay selection methods have been recently
introduced in \cite{Jing,Clarke,Ding,Song,Talwar}. Among these
approaches, a greedy algorithm is an effective way to approach the
global optimal solution. Greedy algorithms have been widely applied
in sparse approximation \cite{Tropp}, internet routing \cite{Flury}
and arithmetic coding \cite{Jia}. In cooperative relaying systems,
greedy algorithms are used in \cite{Ding,Song,lowbit} to search for
the best relay combination. However, with insufficient number of
combinations considered in the selection process, a significant
performance loss is experience as compared to an exhaustive search.

The objective of this paper is to propose a cross-layer design
strategy that jointly considers the optimization of a low-complexity detection and a relay selection
algorithm for ad hoc and sensor networks that employ DS-CDMA
systems. Cross-layer designs that integrate different layers of the network have been employed in prior
work \cite{RCDL2,Chen} to guarantee the quality of service and help increase the
capacity, reliability and coverage of networks. However, there are very few works in the literature involving MUD techniques
(operated in the lower physical layer)
with relay selection (conducted in the data and link layer) in cooperative relaying systems. In
\cite{Venturino,Cao}, an MMSE-MUD technique has been applied
to cooperative systems, the results reveal that the transmissions become
more resistant to MAI and obtain a significant performance gain when
compared with a single direct transmission. However, extra
complexity is introduced, as matrix inversions are required
when an MMSE filter is employed.

In this work, we devise a low-cost greedy list-based successive
interference cancellation (GL-SIC) strategy with RAKE receivers as
the front-end that can approach the maximum likelihood detector
performance. Other more advanced front-ends
\cite{mcg,cgb,ccmmwf,wlmwf,ifir,aifir1,aifir2,jio,stjio,smjio,jidf,sjidf,ccmrls,stbcccm,baidf}
could be easily incorporated. Unlike prior art, the proposed GL-SIC
algorithm exploits the Euclidean distance between users of interest
and their nearest constellation points, with multiple ordering at
each stage, we are able to build all possible lists of tentative
decisions for each user via a greedy-like approach. We also present
a low-cost multi-relay selection algorithm based on greedy
techniques that can approach the performance of an exhaustive
search. In the proposed greedy algorithm, a selection rule is
employed and the process is separated into several stages.
%This process results a low-cost strategy to approach the performance of
%an exhaustive search.At each stage, a limited
%number of relay combinations is examined and compared,
%resulting in a low-cost strategy to approach the performance of
%an exhaustive search.
A cross-layer design technique that brings
together the proposed GL-SIC algorithm and the greedy relay
selection is then considered and evaluated by computer simulations.

The rest of this paper is organized as follows. In Section 2, the
system model is described. In Section 3, the GL-SIC multiuser
detection method is presented. In Section 4, the relay selection
strategy is proposed. In Section 5, simulation results are
presented and discussed. Finally, conclusions are drawn in Section
6.
\vspace{-0.6cm}

\section{Cooperative DS-CDMA system model}
\vspace{-0.35cm}
%\begin{figure}[!htb]
%\begin{center}
%\def\epsfsize#1#2{0.75\columnwidth}
%\epsfbox{Fig1} \caption{\footnotesize%\vspace{-2.85em}
%Uplink of a cooperative DS-CDMA system.} \vskip -5pt \label{fig1}
%\end{center}
%\end{figure}

We consider the uplink of a synchronous DS-CDMA system with $K$
users $(k_1,k_2,...k_K)$, $L$ relays $(l_1,l_2,...l_L)$, $N$ chips per
symbol and $L_p$ $(L_p<N)$ propagation paths for each link. The system
is equipped with a DF protocol at each relay and we assume that the
transmit data are organized in packets comprising $P$ symbols. The
received signals are filtered by a matched filter, sampled at chip
rate to obtain sufficient statistics and organized into $M \times1$
vectors $\textbf{y}_{sd}$, $\textbf{y}_{sr}$ and $\textbf{y}_{rd}$,
which represent the signals received from the sources (users) to the
destination, the sources to the relays and the relays to the
destination, respectively. The proposed algorithms for interference
mitigation and relay selection are employed at the relays and at the
destination. %As shown in Fig. \ref{fig1},
The received signal at the destination comprises the data transmitted during two phases that are
jointly processed at the destination. Therefore, the received signal is described by a
$2M\times1$ vector formed by stacking the received signals
from the relays and the sources as given by
\begin{equation}
\begin{split}
\hspace{-0.5em} \left[\hspace{-0.5em}\begin{array}{l}
  \textbf{y}_{sd}\\
  \textbf{y}_{rd}\\
\end{array} \hspace{-0.5em} \right] & = \left[\hspace{-0.5em} \begin{array}{l}
  \sum\limits_{k=1}^K  a_{sd}^k\textbf{S}_k\textbf{h}_{sd,k}b_k \\
  \sum\limits_{l=1}^{L}\sum\limits_{k=1}^K a_{r_ld}^k\textbf{S}_k\textbf{h}_{r_ld,k}\hat{b}_{r_ld,k}\\
\end{array}\hspace{-0.5em} \right] + \left[ \hspace{-0.5em} \begin{array}{l}
  \textbf{n}_{sd}\\
  \textbf{n}_{rd}\\
  \end{array}\right], \label{equation1}
\end{split}
\end{equation}
where $M=N+L_p-1$, $b_k\in\{+1,-1\}$ correspond to the transmitted
symbols, $a_{sd}^k$ and $a_{r_ld}^k$ represent the $k$-th user's amplitude from
the source to the destination and from the $l$-th relay to the destination.
The $M \times L_p$ matrix $\textbf{S}_k$ contains the signature sequence of each
user shifted down by one position at each column that forms
\begin{equation}
\textbf{S}_k = \left[\begin{array}{c c c }
s_{k}(1) &  & {\bf 0} \\
\vdots & \ddots & s_{k}(1)  \\
s_{k}(N) &  & \vdots \\
{\bf 0} & \ddots & s_{k}(N)  \\
 \end{array}\right],
\end{equation}
where $\textbf{s}_k=[s_k(1),s_k(2),...s_k(N)]^T$ is the signature
sequence for user $k$. The vectors
$\textbf{h}_{sd,k}$, $\textbf{h}_{r_ld,k}$ are the $L_p\times1$
channel vectors for user $k$ from the source to the destination and the $l$-th
relay to the destination. The $M\times1$ noise vectors
$\textbf{n}_{sd}$ and $\textbf{n}_{rd}$ contain samples of zero
mean complex Gaussian noise with variance $\sigma^2$, $\hat{b}_{r_ld,k}$
is the decoded symbol for user $k$ at the output of relay $l$ after using the DF protocol.
The received signal in (\ref{equation1}) can then be described by
\begin{equation}
\textbf{y}_d(i)=\sum\limits_{k=1}^K \textbf{C}_k
\textbf{H}_k(i)\textbf{A}_k(i)\textbf{B}_k(i)+\textbf{n}(i),
\end{equation}
where $i$ denotes the time instant corresponding to one symbol in the
transmitted packet and its received and relayed copies. $\textbf{C}_k$
is a $2M\times(L+1)L_p$ matrix comprising shifted versions of $\textbf{S}_k$ as given by
\begin{equation}
\textbf{C}_k = \left[\begin{array}{c c c c}
\textbf{S}_{k} & {\bf 0} & \ldots & {\bf 0} \\
{\bf 0} & \ \ \textbf{S}_{k} & \ldots & \ \ \textbf{S}_{k}\\
 \end{array}\right],
\end{equation}
$\textbf{H}_k(i)$ represents a $(L+1)L_p \times (L+1)$ channel matrix
between the sources and the destination and the relays and the
destination links. $\textbf{A}_k(i)$ is a $(L+1)\times(L+1)$
diagonal matrix of amplitudes for user $k$. The matrix
$\textbf{B}_k(i)=[b_k,\hat{b}_{r_1d,k},\hat{b}_{r_2d,k},...\hat{b}_{r_Ld,k}]^T$
is a $(L+1)\times1$ vector for user $k$ that contains the transmitted symbol at
the source and the detected symbols at the output of each relay, and
$\textbf{n}(i)$ is a $2M\times1$ noise vector.
\vspace{-0.35cm}

\section{Proposed GL-SIC multiuser detection}
\vspace{-0.35cm}
In this section, we propose the GL-SIC multiuser detection method
that can be applied at both the relays and the destination in the uplink
of a cooperative system. The GL-SIC detector uses the RAKE
receiver as the front-end, which reduces computational complexity by
avoiding the matrix inversion required when MMSE filters are
applied. The GL-SIC detector exploits the Euclidean distance between
the users of interest and their nearest constellation points, with multiple
ordering at each stage, we are able to build all possible lists of tentative decisions
for each user. When seeking appropriate candidates, a greedy-like
technique is performed to build each list and all possible
estimates within the list are examined when unreliable users are
detected. Unlike prior work which employs the concept of Euclidean distance
with multiple feedback SIC (MF-SIC) \cite{Li1}, the proposed GL-SIC
jointly considers multiple numbers of users, constellation constraints and
re-ordering at each detection stage to obtain an improvement in
detection performance.
\vspace{-0.35cm}

\subsection{The proposed GL-SIC design}

\vspace{-0.35cm}
\begin{figure}[!htb]
\vspace{-0.7em}
\begin{center}
\def\epsfsize#1#2{1\columnwidth}
\epsfbox{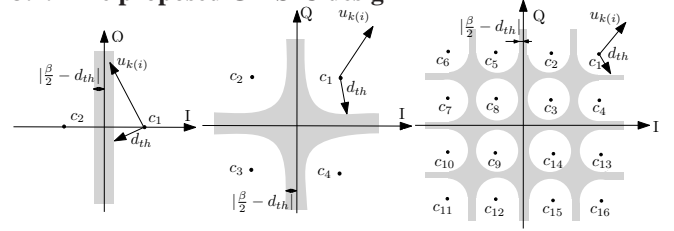} \vspace{-2.5em}
\caption{\footnotesize
The reliability check for soft estimates in BPSK, QPSK and 16 QAM constellations.}
\vspace{-1.5em}
\label{fig2}
\end{center}
\end{figure}

In the following, we describe the process for initially detecting
$n$ users described by the indices $k_1, k_2,...,k_n$ at the first
stage. Other users can be obtained accordingly. As shown by Fig.
\ref{fig2}, $\beta$ is the distance between two nearest
constellation points, $d_{th}$ is the threshold. The soft output of
the RAKE receiver for user $k$ is then obtained by
\begin{equation}
u_{k}(i)=\textbf{w}_{k}^{H}\textbf{y}_{sr_l}(i),
\end{equation}
where $\textbf{y}_{sr_l}(i)$ represents the received signal
from the source to the $l$-th relay, $u_{k}(i)$ stands for the soft output of the $i$-th symbol for
user $k$ and $\textbf{w}_{k}^{H}$ denotes the RAKE receiver that
corresponds to a filter matched to the effective signature at the
receiver. After that, we order all users into a
decreasing power level and organize them into a vector $\textbf{t}_a$. We pick the first
$n$ entries $[\textbf{t}_a(1), \textbf{t}_a(2),...,\textbf{t}_a(n)]$ which denote users $k_1,
k_2,...,k_n$, the reliability of each of the $n$ users is examined by the corresponding Euclidean distance
between the desired user and its nearest constellation point $c$.\\\vspace{-1.2em}
\\
\textbf{Decision reliable}:\\
If all $n$ users satisfy the following condition, they are determined as reliable, namely
\begin{equation}
u_{\textbf{t}_a(t)}(i)\notin \textbf{C}_{\rm grey},\ \ \ {\rm for} \ t\in [1,2,...,n],
\end{equation}
these soft estimates will be applied to a slicer $Q(\cdot)$ as described by
\begin{equation}
\hat{b}_{\textbf{t}_a(t)}(i)=Q(u_{\textbf{t}_a(t)}(i)),\ {\rm for}\ t\in [1,2,...,n],
\end{equation}
where $\hat{b}_{\textbf{t}_a(t)}(i)$ denotes the detected symbol for the $\textbf{t}_a(t)$-th user, $\textbf{C}_{\rm grey}$ is the shadowed area in Fig. \ref{fig2}, it should be noted that the shadowed region would spread along both the vertical and horizontal directions. After that, the cancellation is then performed in the same way as a conventional SIC where we mitigate the impact of MAI brought by these users:
\begin{equation}
\textbf{y}_{sr_l,s+1}(i)=\textbf{y}_{sr_l,s}(i)-\sum\limits_{t=1}^{n}\textbf{H}_{sr_l,\textbf{t}_a(t)}(i)
\hat{b}_{\textbf{t}_a(t)}(i), \label{equation2}
\end{equation}
where $\textbf{H}_{sr_l,\textbf{t}_a(t)}(i) = a_{sr_l}^{\textbf{t}_a(t)}\textbf{S}_{\textbf{t}_a(t)}(i)\textbf{h}_{sr_l,\textbf{t}_a(t)}(i)$
stands for the desired user's channel matrix associated with the link between the
source and the $l$-th relay, $\textbf{y}_{sr_l,s}$ is the received
signal from the source to the $l$-th relay at the $s$-th $(s=1,2,...,K/n)$ cancellation stage.
The above process is repeated subsequently with another $n$ users being selected from the remaining users at each following stage,
and this algorithm changes to the unreliable mode when unreliable users are detected. Additionally, since the interference
created by the previous users with the strongest power has been mitigated, improved estimates are obtained by reordering the remaining users.\\\vspace{-1.2em}
\\
\textbf{Decision unreliable}:\\
(a). If part of the $n$ users are determined as reliable, while others
are considered as unreliable, we have
\begin{equation}
u_{\textbf{t}_p(t)}(i) \notin \textbf{C}_{\rm grey},\ \ \ {\rm for}\ t\in [1,2,...,n_p], \label{equation3}
\end{equation}
\begin{equation}
u_{\textbf{t}_q(t)}(i) \in \textbf{C}_{\rm grey},\ \ \ {\rm for}\ t\in [1,2,...,n_q], \label{equation4}\vspace{-0.8em}
\end{equation}
where $\textbf{t}_p$ is a $1 \times n_p$ vector that
contains $n_p$ reliable users and $\textbf{t}_q$ is a $1 \times n_q$ vector
that includes $n_q$ unreliable users, subject to $\textbf{t}_p\cap \textbf{t}_q=\varnothing$ and $ \textbf{t}_p\cup \textbf{t}_q=[1,2,...n]$ with $n_p+n_q=n$. Consequently, the $n_p$ reliable users are applied to the slicer $Q(\cdot)$ directly and the $n_q$ unreliable ones are examined in terms of all possible constellation values $c^m$ $(m=1,2,...,N_c)$ from the $1\times N_c$ constellation points set $\textbf{C}\subseteq \textsl{F}$, where $\textsl{F}$ is a subset of the complex field and $N_c$ is determined by the modulation type. The detected symbols are given by
\begin{equation}
\hat{b}_{\textbf{t}_p(t)}(i)=Q(u_{\textbf{t}_p(t)}(i)),
{\rm for}\ t\in [1,2,...,n_p],
\end{equation}
\begin{equation}
\hat{b}_{\textbf{t}_q(t)}(i)=c^m,\ \
{\rm for}\ t\in [1,2,...,n_q],\vspace{-0.7em}
\end{equation}
At this point, $N_c^{n_q}$ combinations of candidates for $n_q$ users are generated. The detection tree is then split into $N_c^{n_q}$ branches. After this processing, (\ref{equation2}) is applied with its corresponding combination to ensure the influence brought by the $n$ users is mitigated. Following that, $N_c^{n_q}$ updated $\textbf{y}_{sr_l}(i)$ are generated, we reorder the remaining users at each cancellation stage and compute a conventional SIC with RAKE receivers for each branch. The following $K\times1$ different ordered candidate detection lists are then produced
\begin{equation}
\textbf{b}^j(i)=[\textbf{s}_{\textrm{pre}}(i), \ \ \textbf{s}_p(i),\ \ \textbf{s}^j_q(i),\ \ \textbf{s}^j_{\textrm{next}}(i)]^T, \ j=1,2,...,N_c^{n_q},
\end{equation}
where $\textbf{s}_{\textrm{pre}}(i)=[\hat{b}_{\textbf{t}_a(1)}(i),\hat{b}_{\textbf{t}_a(2)}(i),...]^T$ stands for
the previous stages detected reliable symbols, $\textbf{s}_p(i)=[\hat{b}_{\textbf{t}_p(1)}(i), \hat{b}_{\textbf{t}_p(2)}(i)\\,...,\hat{b}_{\textbf{t}_p(n_p)}(i)]^T$ is a $n_p\times 1$ vector that denotes the current stage reliable symbols detected directly from slicer $Q(\cdot)$ when (\ref{equation3}) occurs, $\textbf{s}^j_q(i)=[c_{\textbf{t}_q(1)}^{m},c_{\textbf{t}_q(2)}^{m},...,c_{\textbf{t}_q(n_q)}^{m}]^T, j=1,2,...,N_c^{n_q}$
is a $n_q\times 1$ vector that contains the detected symbols deemed unreliable at the current stage
under the situation of (\ref{equation4}), each entry of this vector is selected randomly from the constellation
point set $\textbf{C}$ and all possible $N_c^{n_q}$ combinations need to be considered and examined.
$\textbf{s}^j_{\textrm{next}}(i)=[...,\hat{b}_{\textbf{t}'(1)}^{\textbf{s}^j_q}(i),...,\hat{b}_{\textbf{t}'(n)}^{\textbf{s}^j_q}(i)]^T$
 includes the corresponding detected symbols in the following stages after the $j$-th combination of $\textbf{s}_q(i)$ is allocated to the unreliable user vector $\textbf{t}_q$, and $\textbf{t}'$ is a $n \times 1$ vector that contains the users from the last stage.\\\vspace{-1.2em}
\\
(b). If all $n$ users are considered as unreliable:
\begin{equation}
u_{\textbf{t}_b(t)}(i) \in \textbf{C}_{\rm grey},\ \ \ {\rm for}\ t\in [1,2,...,n],\label{equation5}
\end{equation}
where $\textbf{t}_b=[1,2,...,n]$, then all $n$ unreliable users can assume the values in $\textbf{C}$.
In this case, the detection tree will be split into $N_c^{n}$ branches to produce
\begin{equation}
\hat{b}_{\textbf{t}_b(t)}(i)=c^m, \ {\rm for}\ t\in [1,2,...,n],
\end{equation}
Similarly, (\ref{equation2}) is then applied and a conventional SIC with different orderings at each cancellation stage
is performed via each branch. Since all possible constellation values are tested for all unreliable users, we have the candidate lists:
\begin{equation}
\textbf{b}^j(i)=[\textbf{s}_{\textrm{pre}}(i),\ \textbf{s}_b^j(i),\ \ \textbf{s}^j_{\textrm{next}}(i)]^T, \ j=1,2,...,N_c^n,
\end{equation}
where $\textbf{s}_{\textrm{pre}}(i)=[\hat{b}_{\textbf{t}_a(1)}(i),\hat{b}_{\textbf{t}_a(2)}(i),...]^T$ are the reliable symbols that are detected from previous stages, $\textbf{s}_b^j(i)=[c_{\textbf{t}_b(1)}^{m},c_{\textbf{t}_b(2)}^{m},...,c_{\textbf{t}_b(n)}^{m}]^T, j=1,2,...,N_c^n$ is a $n\times1$ vector that represents $n$ number of users which are regarded as unreliable at the current stage as shown by (\ref{equation5}), each entry of $\textbf{s}_b^j$ is selected randomly from the constellation point set $\textbf{C}$. The vector $\textbf{s}^j_{\textrm{next}}(i)=[...,\hat{b}_{\textbf{t}'(1)}^{\textbf{s}_b^j}(i),...,\hat{b}_{\textbf{t}'(n)}^{\textbf{s}_b^j}(i)]^T$ contains the corresponding detected symbols in the following stages after the $j$-th combination of $\textbf{s}_b(i)$ is allocated to all unreliable users.
After the candidates are generated, lists are built for each group of users, and the ML rule is used at the end of
the detection procedure to choose the best candidate list as described by
\begin{equation}
\textbf{b}^{\textrm{best}}(i)= \min _{\substack{1\leq j\leq m, \textrm{where}\\m=N_c^{n_q} \textrm{or}\ N_c^n}}\parallel \textbf{y}_{sr_l}(i)-\textbf{H}_{sr_l}(i)\textbf{b}^{j}(i)\parallel^2.
\end{equation}
\vspace{-0.55cm}

\subsection{GL-SIC with multi-branch processing}
\vspace{-0.35cm}
The multiple branch (MB) structure employs multiple parallel processing branches which can help to regenerate extra detection diversity has been discussed carefully in \cite{RCDL1,Li2}. Inspired by that, an extended structure that incorporates the proposed GL-SIC with multi-branch is achieved. By changing the obtained best detection order for $\textbf{b}^{\textrm{best}}$ with indices $\textbf{O}=[1,2,...,K]$ into a group of different detection sequences, we are able to gain this parallel structure with each branch shares a different detection order. Since it is not practical to test all $L_b=K!$ possibilities due to the high complexity, a reduced number of branches can be tested, with each index number in $\textbf{O}_{l_b}$
being the corresponding index number in $\textbf{O}$ cyclically shifted to right by one position, namely:
$\textbf{O}_{l_1}=[K,1,2,...,K-2,K-1], \textbf{O}_{l_2}=[K-1,K,1,...,K-3,K-2],..., \textbf{O}_{l_{K-1}}=[2,3,4,...,K,1]$
and $\textbf{O}_{l_K}=[K,K-1,...,1]$(reversed order).
After that, each of the $K$ parallel branches computes a GL-SIC algorithm with its corresponding order, respectively.

After obtaining $K+1$ ($\textbf{O}$ included) different candidate lists according to each of the branch, a modified ML rule is applied with the following steps:

\begin{enumerate}
\item Obtain the best detection candidate branch $\textbf{b}^{O_{l_{\rm base}}}(i)$ among all $K+1$ parallel branches according to the ML rule
\begin{equation}
\textbf{b}^{O_{l_{\rm base}}}(i)= \min _{\substack{0\leq b\leq K}}\parallel \textbf{y}_{sr_l}(i)-\textbf{H}_{sr_l}\textbf{b}^{O_{l_b}}(i)\parallel^2.\label{equation6}
\end{equation}
\item Re-examine the detected value for user $k$ $(k=1,2,...,K)$ by fixing the detected results of all other unexamined users in $\textbf{b}^{O_{l_{\rm base}}}(i)$, then replace the $k$-th user's detection result $\hat b_{k}$ in $\textbf{b}^{O_{l_{\rm base}}}(i)$ by its corresponding detected values from all other $K$ branches $\textbf{b}^{O_{l_b}}(i)$ $(b=0,1,...,K, l_b\neq l_{\rm base}, \textbf{O}=\textbf{O}_{l_0})$ with the same index, the combination with the minimum Euclidean distance is selected through (\ref{equation6}) and the corresponding improved estimate of user $k$ is saved and kept.
\item The same process is then automatically repeated with the next
      user in $\textbf{b}^{O_{l_{\rm base}}}(i)$ until all users in $\textbf{b}^{O_{l_{\rm base}}}(i)$ are examined.
\end{enumerate}

As each combination selection seeks an improvement based on the previous choice, a better candidate list $\textbf{b}^{\textrm{opt}}(i)$ can be generated. This is performed at all the relays and the destination using the RAKE receiver structure.
\vspace{-0.45cm}

\section{proposed greedy multi-relay selection method}
\vspace{-0.35cm}
In this section, a greedy multi-relay selection method is introduced. For this problem, the best relay combination is obtained through an exhaustive search of all possible subsets of relays. However, the major problem that prevents us from applying an exhaustive search in practical communications is its high computational complexity. With $L$ relays involved in the transmission, an exponential complexity of $2^L-1$ is experienced. This fact motivates us to seek other alternative methods. The standard greedy algorithm can be used in the selection process by cancelling the poorest relay-destination link stage by stage, however this method can approach only a local optimum. The proposed greedy multi-relay selection method can go through a reduced number of relay combinations and approach the best one based on previous decisions. In the proposed relay selection, the signal to interference and noise ratio (SINR) is used as the criterion to determine the optimum relay set. The expression of the SINR is given by
\begin{equation}
 {\rm SINR_q} =\frac{E[|\textbf{w}_q^H\textbf{r}|^2]}{E[|\boldsymbol\eta|^2]+\textbf{n}},
\end{equation}
where $\textbf{w}_q$ denotes the RAKE receiver for user $q$, $E[|\boldsymbol\eta|^2]=E[|\sum\limits_{\substack{k=1\\k\neq q}}^K\textbf{H}_kb_k|^2]$ is the interference brought by all other users, $\textbf{n}$ is the noise. For the RAKE receiver, the SINR is derived by
\begin{equation}
 {\rm SINR_q} =\frac{\textbf{h}_q^H\textbf{H}\textbf{H}^H\textbf{h}_q} {\textmd{trace}(\textbf{H}_\eta \textbf{H}_\eta^H)+\textbf{h}_q^H\sigma_N^2\textbf{h}_q},
\end{equation}
where $\textbf{h}_q$ is the channel vector for user $q$, $\textbf{H}$ is the channel matrix for all users, $\textbf{H}_\eta$ represents the channel matrix of all other users except user $q$. It should be mentioned that in various relay combinations, the channel vector $\textbf{h}_q$ for user $q$ $(q=1,2,...,K)$ is different as different relay-destination links are involved, $\sigma_N^2$ is the noise variance. This problem thus can be cast as the following optimization:
\begin{equation}
 {\rm SINR_{\Omega_{best}}}= max \ \ \{min({\rm SINR_{\Omega_{r(q)}}}), q=1,...,K\},
\end{equation}
where $\Omega_{r}$ denotes all possible combination sets $(r \leq L(L+1)/2)$ of any number of selected relays, ${\rm SINR_{\Omega_{r(q)}}}$ represents the SINR for user $q$ in set $\Omega_r$, min $({\rm SINR_{\Omega_{r(q)}}}) = {\rm SINR_{\Omega_r}}$ stands for the SINR for relay set $\Omega_r$ and
$\Omega_{\rm best}$ is the best relay set that provides the highest SINR.
\vspace{-0.5cm}

\subsection{Standard greedy relay selection algorithm}
\vspace{-0.35cm}
The standard greedy relay selection method is operated in stages by cancelling the single relay according to the channel condition, as the channel path power is given by
\begin{equation}
 P_{h_{r_ld}}=\textbf{h}_{r_ld}^H\textbf{h}_{r_ld},
\end{equation}
where $\textbf{h}_{r_ld}$ is the channel vector between the $l$-th relay and the destination. The selection begins with all $L$ relays participating in the transmission, and the initial SINR is determined when all relays are involved in the transmission. For the second stage, we cancel the poorest relay-destination link and calculate the current SINR for the remaining $L-1$ relays, as compared with the previous SINR, if
\begin{equation}
{\rm SINR_{cur}}>{\rm SINR_{pre}}, \label{equation7}
\end{equation}
we update the previous SINR as
\begin{equation}
{\rm SINR_{pre}} = {\rm SINR_{cur}}, \label{equation8}
\end{equation}
and move to the third stage, where we remove the current poorest link and repeat the above process. The algorithm will stop when ${\rm SINR_{cur}}<{\rm SINR_{pre}}$ or when there is only one relay left. The whole process can be performed once before each packet transmission.
\vspace{-0.5cm}

\subsection{Proposed greedy relay selection algorithm}
\vspace{-0.35cm}
In order to improve the performance of the existing algorithm, we have modified the above standard greedy algorithm. The proposed method differs from the standard technique as we drop each of the relays in turns rather than drop them based on the channel condition at each stage. The algorithm can be summarized as:
\begin{enumerate}
\item Initially, a set $\Omega_A$ that includes all $L$ relays is generated and its corresponding SINR is calculated, denoted by ${\rm SINR_{pre}}$.
\item For the second stage, we calculate the SINR for $L$ combination sets with each dropping one of the relays from $\Omega_A$. After that, we pick the combination set with the highest SINR for this stage, recorded as ${\rm SINR_{cur}}$.
\item Compare ${\rm SINR_{cur}}$ with the previous stage ${\rm SINR_{pre}}$, if (\ref{equation7}) is true, we save this corresponding relay combination as $\Omega_{\textrm{cur}}$ at this stage. Meanwhile, we update the ${\rm SINR_{pre}}$ as in (\ref{equation8}).
\item After moving to the third stage, we drop relays in turn again from $\Omega_{\textrm{cur}}$ obtained in stage two. $L-1$ new combination sets are generated, we then select the set with the highest SINR and repeat the above process in the following stages until either ${\rm SINR_{cur}}<{\rm SINR_{pre}}$ or there is only one relay left.
\end{enumerate}
This new greedy selection method considers the combination effect of the channel condition, which implies additional useful sets are examined. When compared with the standard greedy relay selection method, the previous stage decision is more accurate and the global optimum can be approached more closely. Similarly, the whole process is performed only once before each packet. Meanwhile, its complexity is less than $L(L+1)/2$, much lower than the exhaustive search.\vspace{-1.2em}
\begin{table}[!htp]\scriptsize
\footnotesize
\centering\caption{The proposed greedy multi-relay selection algorithm}
\begin{tabular}{l}
\hline
$\Omega_A=[1,2,3,...L]$\% $\Omega_A$ denotes the set when all relays are involved\\
${\rm SINR_{\Omega_A}}=\textrm{min}({\rm SINR_{\Omega_{A(q)}}}), q=1,2,...K$\\
${\rm SINR_{pre}}={\rm SINR_{\Omega_A}}$\\
\textbf{for} stage =1 \textbf{to} $L-1$\\ \ \ \ \ \
\textbf{for} $r$=1 \textbf{to} $L+1$-stage\\ \ \ \ \ \ \ \ \ \
$\Omega_r=\Omega_A-\Omega_{A(r)}$\% drop each of the relays in turns\\ \ \ \ \ \ \ \ \ \
${\rm SINR_{\Omega_r}}=\textrm{min}({\rm SINR_{\Omega_r(q)}}), q=1,2,...,K$\\\ \ \ \ \
\textbf{end for}\\ \ \ \ \ \
${\rm SINR_{\textrm{cur}}}=\textrm{max}({\rm SINR_{\Omega_r}})$\\ \ \ \ \ \
$\Omega_{\textrm{cur}}=\Omega_{{\rm SINR_{cur}}}$\\ \ \ \ \ \
\textbf{if} ${\rm SINR_{cur}}>{\rm SINR_{pre}}$ \textbf{and} $\rm{length}(\Omega_{\textrm{cur}})>1$\\ \ \ \ \ \ \ \ \ \
$\Omega_A=\Omega_{\textrm{cur}}$\\ \ \ \ \ \ \ \ \
${\rm SINR_{pre}}={\rm SINR_{cur}}$\\ \ \ \ \ \
\textbf{else}\\ \ \ \ \ \ \ \ \ \ \
\textbf{break}\\ \ \ \ \
\textbf{end if}\\
\textbf{end for}\\
\hline
\end{tabular}
\vspace{-1.2em}
\end{table}
\vspace{-0.25cm}
\section{simulations}
\vspace{-0.35cm}
In this section, a simulation study of the proposed GL-SIC multiuser detection strategy with a RAKE receiver and the low cost greedy multi-relay selection method is carried out. The DS-CDMA network uses randomly generated spreading codes of length $N=16$ and $N=32$, it also employs $L_p=3$ independent paths with the power profile $[0\rm dB,-3\rm dB,-6\rm dB]$ for each source-relay, source-destination and relay-destination link, their corresponding channel coefficients $h_{sr_l}^{l_p}$, $h_{sd}^{l_p}$ and $h_{r_ld}^{l_p} (l_p=1,2,...,L_p)$ are taken as uniformly random variables and normalized to ensure the total power is unity. We assume perfectly known channels at the receiver. Equal power allocation with normalization is assumed to guarantee no extra power is introduced during the transmission.  We consider packets with 1000 BPSK symbols and average the curves over 300 trials. For the purpose of simplicity, $d_{th}=0.25$ and BPSK modulation technique are applied in the following simulations. Furthermore, in the GL-SIC algorithm, $n=2$ users are considered at each stage. The performance is evaluated in both non-cooperative and cooperative schemes.
\vspace{-0.3cm}
\begin{figure}[!htb]
\begin{center}
\def\epsfsize#1#2{0.8\columnwidth}
\epsfbox{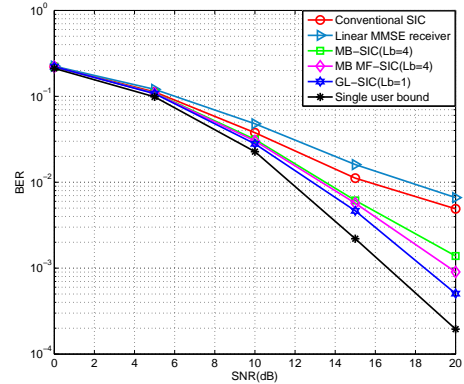} \vspace{-1.5em}\caption{\footnotesize
Non-cooperative system with N=32, 20 users over Rayleigh fading channel}
\label{fig3}
\end{center}
\vspace{-1.5em}
\end{figure}

The first example shown in Fig. \ref{fig3} is the proposed GL-SIC interference suppression technique with 20 users that only takes into account the source-destination link. The conventional SIC detector is the standard SIC with RAKE receivers employed at each stage and the Multi-branch Multi-feedback (MB-MF) SIC detection algorithm mentioned in \cite{Li1} is presented here for comparison purposes. We also produced the simulation results for the multi-branch SIC (MB-SIC) where four parallel branches ($L_b=4$) with different detection orders are applied. Simulation results reveal that our proposed single branch GL-SIC significantly outperforms the linear MMSE receiver, the conventional SIC and exceeds the performance of MB-SIC with $L_b=4$ and MB MF-SIC with $L_b=4$ for the same BER performance.
\vspace{-0.25cm}
\begin{figure}[!htb]
\centerline{
\includegraphics[width=0.18\textwidth]{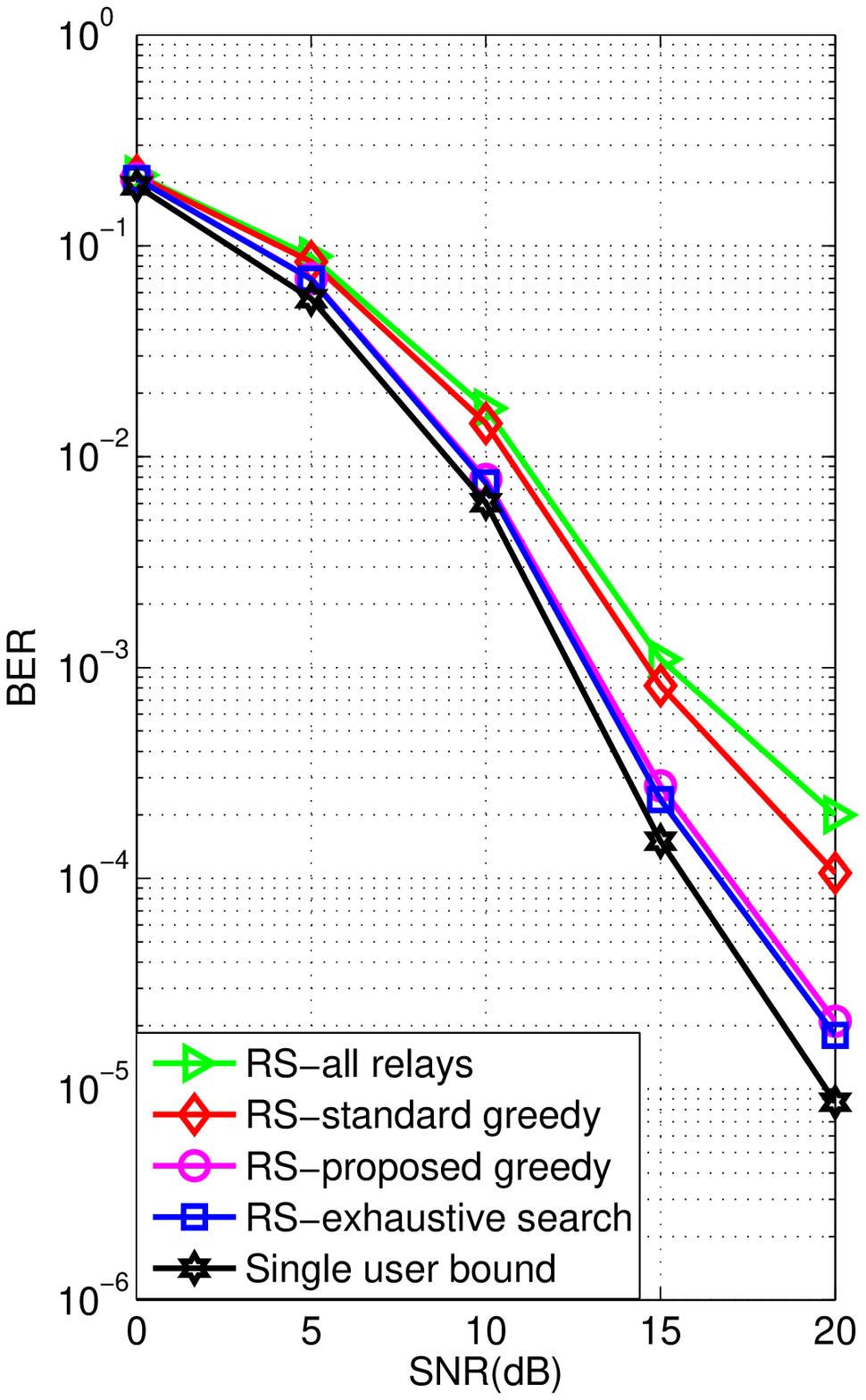}
\includegraphics[width=0.18\textwidth]{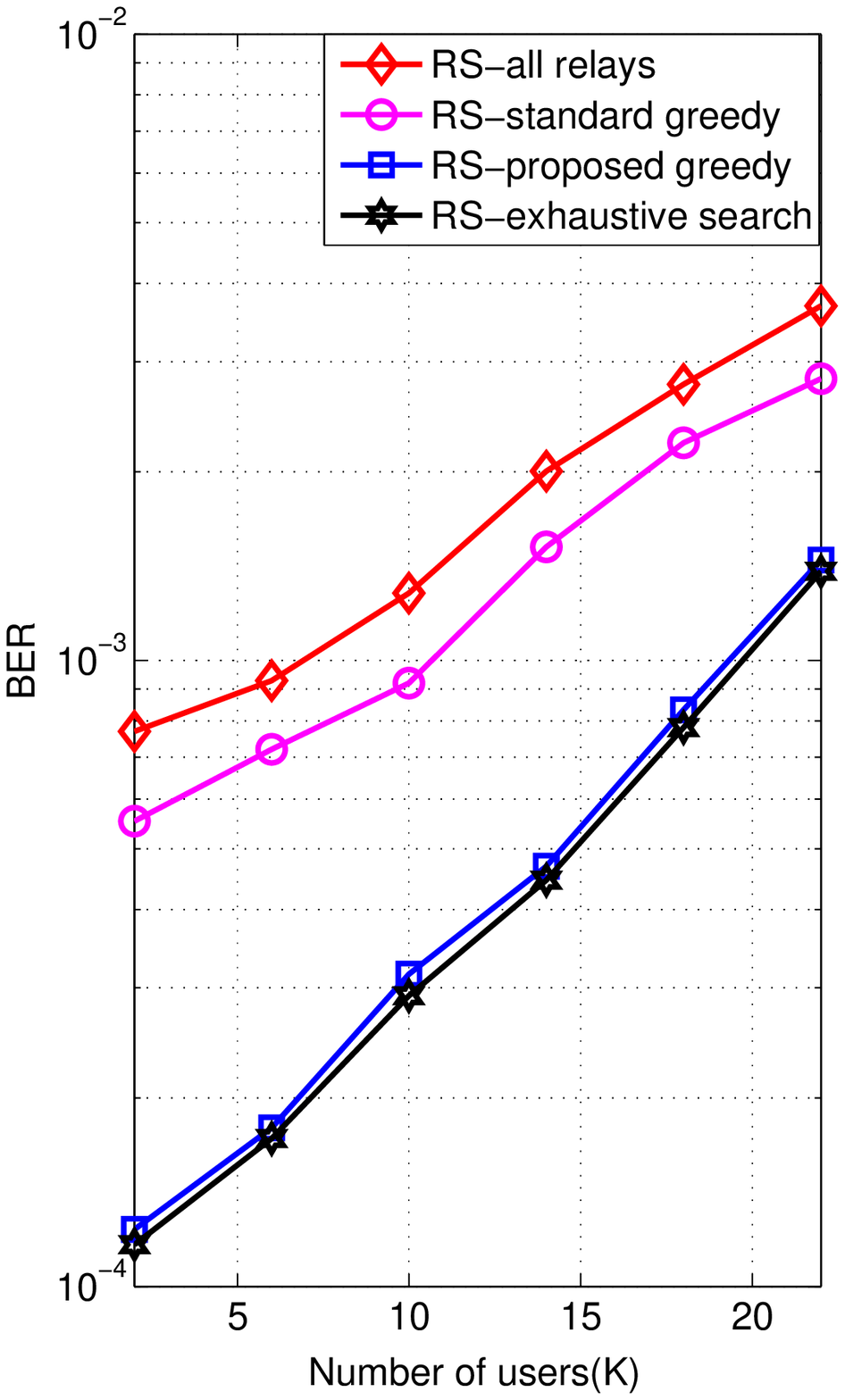}
}\vspace{-0.55em}\caption{\footnotesize
a)\ BER versus SNR for uplink cooperative system (N=16)\ \ \
b)\ BER versus number of users for uplink cooperative system (N=16)}
\label{fig4}
\vspace{-1em}
\end{figure}

The second example denotes the cooperative system where relays assist to process and forward the information. Fig. \ref{fig4}(a) shows the BER versus SNR plot for the different multi-relay selection strategies, where we apply the GL-SIC $(L_b=1)$ detection scheme at both the relays and the destination in an uplink cooperative scenario with 10 users and 6 relays. The performance bound for exhaustive search is presented here for comparison, where it examines all possible relay combinations and picks the best one with the highest SINR. From the results, it can be seen that with relay selection, the BER performance substantially improves. Furthermore, the BER performance curve of our proposed algorithm outperforms the standard greedy algorithm and approaches to the same level of the exhaustive search, whilst keeping the complexity reasonably low for practical utilization.

The algorithms are then assessed in terms of the BER versus number of users in Fig. \ref{fig4}(b) with a fixed SNR=15dB. Similarly, we apply the GL-SIC $(L_b=1)$ detector at both the relays and the destination. The results indicate that the overall system performance degrades as the number of users increases. It also suggests that our proposed greedy relay selection method is more suitable than the standard greedy relay selection and non-relay selection scenario with a high SNR and a large number of users, as its curve does not change considerably when the number of users increases.
\vspace{-0.3cm}
\begin{figure}[!htb]
\begin{center}
\def\epsfsize#1#2{0.8\columnwidth}
\epsfbox{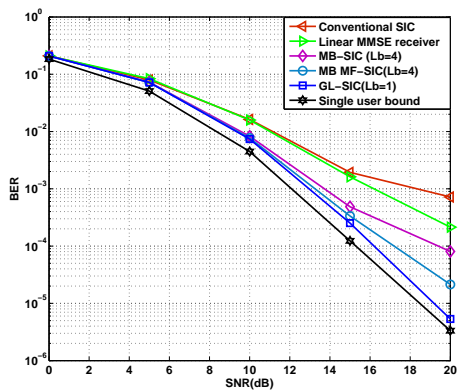} \vspace{-1.55em}\caption{\footnotesize%
BER versus SNR for uplink cooperative system with different filters employed in the relays and the destination (N=16)}
\label{fig5}
\end{center}
\vspace{-1.5em}
\end{figure}

In order to further verify the performance for the proposed cross-layer design, we compare the effect of different detectors with 10 users and 6 relays when this new greedy multi-relay selection algorithm is applied in the system. The results depicted in Fig. \ref{fig5} indicate that the GL-SIC approach allows a more effective reduction of BER and achieves the best performance that is quite close to the single user scenario, followed by the MB MF-SIC detector, the MB-SIC detector, the linear MMSE receiver and the conventional SIC with RAKE receivers employed at each stage.
\vspace{-0.5em}
\vspace{-0.35cm}

\section{conclusion}\vspace{-0.2em}
\vspace{-0.35cm}
A cross-layer design strategy that incorporates the greedy list-based joint successive interference cancellation (GL-SIC) detection technique and a greedy multi-relay selection algorithm for the uplink of cooperative DS-CDMA systems has been presented in this paper. This approach effectively reduces the error propagation generated at the relays, avoiding the poorest relay-destination link while requiring
a low complexity. Simulation results demonstrated that the proposed cross-layer optimization
technique is superior to the existing techniques and can approach the exhaustive search very
closely.
\vspace{-0.35cm}

{\footnotesize
\bibliographystyle{IEEEbib}
\bibliography{reference}}
\end{document}